\documentclass[11pt]{article}
\usepackage{verbatim,amsmath,amssymb}
\usepackage{epsfig,float,color}
\usepackage[font=small,labelfont=bf]{caption}
\usepackage{geometry}
\usepackage{natbib}
\usepackage{hyperref}
\usepackage{setspace}
\usepackage{todonotes}

\definecolor{ured}{rgb}{0.9,0.3,0}
\definecolor{darkblue}{rgb}{0,0,1}
\hypersetup{pdftex=true, colorlinks=true, breaklinks=true, linkcolor=darkblue, menucolor=darkblue, pagecolor=darkblue, citecolor=darkblue, urlcolor=darkblue}

\makeatletter
\newcommand\footnoteref[1]{\protected@xdef\@thefnmark{\ref{#1}}\@footnotemark}
\makeatother

\input{neco_updated.tex}

\begin{document}
\begin{center}
\Large{\bf{Response to David Steigmann's discussion of our paper}}\\
\end{center}

\renewcommand{\thefootnote}{\fnsymbol{footnote}}

\begin{center}
\large{Thang X. Duong$^a$, Mikhail Itskov$^b$ and Roger A. Sauer$^{c,d,e}$\footnote[1]{corresponding author, email: roger.sauer@pg.edu.pl, sauer@aices.rwth-aachen.de}}\\
\vspace{4mm}






\small{$^a$\textit{Institute of Engineering Mechanics \& Structural Analysis, University of the Bundeswehr Munich, Germany}}

\small{$^b$\textit{Department of Continuum Mechanics, RWTH Aachen University, Germany}}

\small{$^c$\textit{Faculty of Civil and Environmental Engineering, Gda{\'n}sk University of Technology, Poland}}

\small{$^d$\textit{Aachen Institute for Advanced Study in Computational Engineering Science (AICES), \\ RWTH Aachen
University, Germany}}

\small{$^e$\textit{Department of Mechanical Engineering, Indian Institute of Technology Guwahati, India}}

\end{center}

\vspace{-6mm}

\renewcommand{\thefootnote}{\arabic{footnote}}

\begin{center}

\small{Published\footnote{This pdf is the personal version of an article whose journal version is available at \href{https://doi.org/10.1177/10812865231224106}{https://journals.sagepub.com}} 
in \textit{Math.~Mech.~Solids}, \href{https://doi.org/10.1177/10812865231224106}{DOI: 10.1177/10812865231224106} \\
Submitted on 16 October 2023; Accepted on 29 November 2023} 

\end{center}

\vspace{-2mm}

We thank Prof.~David Steigmann for his discussion \citep{Steigmann2023}, in particular his proof on the general validity of his internal power expression (see Eq.~\eqref{e:workpairGradientTheoryCM} below) and his expression for the variation of the geodesic curvature. 
His proof allows us to clarify two misleading statements in \cite{shelltextile}, and confirm that its formulation is fully consistent with the formulation of \cite{Steigmann2018}. 
However, some of our original statements made in Remark~4.9 are not wrong: 
The third term in  \eqref{e:workpairGradientTheoryCM} can lead to spurious constitutive models for in-plane bending,
which was our original concern. 

The discussion below follows the notation in \cite{shelltextile}. 
This translates to the notation in Steigmann's discussion as follows: 
$c_\gamma \equiv p_\gamma$,  $c^0_\gamma\equiv M_\gamma$, $\ell^\alpha\equiv l^\alpha$, $\bc_0 \equiv \mM$, $\bc \equiv \mpp$, $\bell \equiv  \ml$, $\kappa_\mrg \equiv\eta_l$, $\kappa^0_\mrg \equiv\eta_L$.

\subsubsection*{Concerning Steigmann's discussion of Remark 4.8}{\label{s:rem48}}

We agree with Steigmann that our statement in Remark 4.8 is incorrect. 
This error stems from our incomplete proof provided in Appendix~5 of \citet{shelltextile}, 
which is based on the incorrect assumption that the variation of the fiber director, $\delta{c}_\alpha$, is not solely expressible through the variation of the metric, $\delta{a}_{\alpha\beta}$. 
Instead, 
\eqb{lll}
\delta{c}_\gamma = \ds \frac{1}{2} c_\gamma\,c^{\alpha\beta}\,\delta{a}_{\alpha\beta}\,,
\label{e:var_c_alpha_s}\eqe
i.e.~$\delta{c}_\alpha$ is solely expressible through $\delta{a}_{\alpha\beta}$.
Eq.~\eqref{e:var_c_alpha_s} follows from the variations $\delta J = J\,\aab\,\delta\auab/2$,  $\delta \lambda = L^{\alpha\beta}\delta\auab/(2\lambda)$, the identity $\aab = \ellab + c^{\alpha\beta}$ that follows from Eq.~(8) in \citet{shelltextile}, and the relation
\eqb{lll}
c_\gamma = J\,\lambda^{-1}\,c^0_\gamma\,,
\label{e:c_alpha_s}\eqe
given in Steigmann's discussion.
Here, $c_\alpha:=\bc\cdot\ba_\alpha$, $J = \det_\mrs\bF$, $\lambda = \norm{\bF\bL}$, $c^0_\alpha:=\bc_0\cdot\bA_\alpha$, $c^{\alpha\beta}:=c^\alpha c^\beta$, $c^\alpha := \bc\cdot\ba^\alpha$, $\ell^{\alpha\beta}:=\ell^\alpha\ell^\beta$, $\ell^\alpha := \bell\cdot\ba^\alpha$, $L^{\alpha\beta}:=L^\alpha L^\beta$ and $L^\alpha := \bL\cdot\bA^\alpha$. 
We thank Steigmann for providing relation \eqref{e:c_alpha_s}. 
We were not aware of this relation in our derivation of Appendix~5 \citep{shelltextile}, since we had used
\eqb{ll}
\delta c_\alpha = - \ell_\alpha^\beta\,\bc\cdot\delta\ba_\beta + \bc\cdot \delta\ba_\alpha
\label{e:var_c_alpha}
\eqe
(see Eq.~(213) of \citet{shelltextile}) instead of Eq.~\eqref{e:var_c_alpha_s}.
We note here that Eq.~\eqref{e:var_c_alpha} is not wrong. 
Instead, it is equivalent to Eq.~\eqref{e:var_c_alpha_s}.
This can be shown by applying the identities $\bc\cdot \delta\ba_\alpha = \delta_\alpha^\beta\,\bc\cdot\delta\ba_\beta$ and $\delta_\alpha^\beta = \ell_\alpha^\beta + c_\alpha^\beta$ (following from Eq.~(8) in \citet{shelltextile}) to \eqref{e:var_c_alpha} and using the symmetry of $c^{\alpha\beta}$.

Relation \eqref{e:c_alpha_s} can also be used to show that the geodesic curvature given in Eq.~(6) of Steigmann's discussion,
\eqb{llll}
\kappa_\mrg \is J\,\lambda^{-3}\,L^{\alpha\beta}\,c^0_{\gamma}\,S_{\alpha\beta}^\gamma -  J\,\lambda^{-3}\,L^{\alpha\beta}\,c^0_{\alpha;\beta}\,, 
\label{e:kappa}
\eqe
is identical to Eq.~(51) in \cite{shelltextile}, i.e.
\eqb{llll}
\kappa_\mrg \is \ell^{\alpha\beta}\,c_{\gamma}\,S_{\alpha\beta}^\gamma + \lambda^{-1}\,c_\alpha\,\ell^\beta\,L^\alpha_{;\beta}\,,
\label{e:kappa2}
\eqe 
This follows from Eq.~\eqref{e:c_alpha_s}, $L^\alpha = \lambda\,\ell^\alpha$ and $L^\alpha\,c^0_{\alpha;\beta} = -c^0_\alpha\,L^\alpha_{;\beta}$ (cf.~Eq.~(38) in \cite{shelltextile}).

With variation \eqref{e:var_c_alpha_s}, we can now continue the derivation in Appendix~5 of \citet{shelltextile}:  
Starting from Eq.~\eqref{e:kappa2} and using Eq.~\eqref{e:var_c_alpha_s} together with Eqs.~(208) and (210) from \citet{shelltextile}, we arrive at the expression
\eqb{lll}
\dot{\kappa}_\mrg = \ds \ellab\,c_\gamma\,\dot{S}^\gamma_{\alpha\beta} +  \frac{1}{2}\,\kappa_\mrg\,(\aab - 3\,\ellab)\,\dot{a}_{\alpha\beta}\,,
\label{e:var_kapp}
\eqe
which is also found in Steigmann's discussion.
That is, the rate of geodesic curvature, $\dot\kappa_\mrg$, is fully expressible in terms of $\dot{a}_{\alpha\beta}$ and $\dot{S}^\gamma_{\alpha\beta}$. 
Since these are symmetric w.r.t.~$\alpha$ and $\beta$, the stress symmetrization employed in the power balance of \cite{Steigmann2018} is indeed general, contrary to what was written in Remark~4.8. 
We thank Prof. Steigmann for pointing out this error.

Eq.~\eqref{e:var_kapp} allows us to confirm that our internal power expression given in Eq.~(107) of \cite{shelltextile},
\eqb{l}
P_{\mathrm{int}} = \ds\frac{1}{2}\int_{\sR_0}\, \tau^{\alpha\beta} \,\dot{a}_{\alpha\beta}\,\dif A + \int_{\sR_0}\Mab_0\,\dot{b}_{\alpha\beta}\,\dif A +  \int_{\sR_0}\bar\mu_0\,\dot\kappa_\mrg\,\dif A\,,
\label{e:workpair1}\eqe
is equivalent to the internal power expression given in \cite{Steigmann2018}, cf.~Eq.~(63), which reads 
\eqb{l}
P_{\mathrm{int}} = \ds\frac{1}{2}\int_{\sR_0}\, \tau^{\alpha\beta} \,\dot{a}_{\alpha\beta}\,\dif A + \int_{\sR_0}\Mab_0\,\dot{b}_{\alpha\beta}\,\dif A +  \int_{\sR_0}\Mbarab_{0\gamma}\,\dot{S}_{\alpha\beta}^\gamma\,\dif A\,,
\label{e:workpairGradientTheoryCM}
\eqe
in our notation.
This follows from inserting Eq.~\eqref{e:var_kapp} into \eqref{e:workpair1} and redefining the stress as
\eqb{l}
\tau^{\alpha\beta} \leftarrow \tau^{\alpha\beta} + \bar{\mu}_0\,\kappa_\mrg\,(\aab - 3\,\ellab)\,.
\label{e:tau}\eqe
Eq.~\eqref{e:workpair1} is also equivalent to 
\eqb{l}
P_{\mathrm{int}} = \ds\frac{1}{2}\int_{\sR_0}\, \tau^{\alpha\beta} \,\dot{a}_{\alpha\beta}\,\dif A + \int_{\sR_0}\Mab_0\,\dot{b}_{\alpha\beta}\,\dif A +  \int_{\sR_0}\Mbarab_0\,\dot{\bar{b}}_{\alpha\beta}\,\dif A\,,
\label{e:workpair3}
\eqe
the internal power expression proposed in \cite{shelltextile} (see Eq.~(113) there).
Therefore the internal power expressions given by \cite{Steigmann2018} (i.e.~Eq.~\eqref{e:workpairGradientTheoryCM}) and our expression in Eq.~\eqref{e:workpair3} are equivalent.
However, the last term in \eqref{e:workpairGradientTheoryCM} can lead to spurious constitutive models for in-plane bending, as is shown below.

\subsubsection*{Concerning Steigmann's discussion of Remark 4.9}{\label{s:rem49}}

Steigmann's discussion of Remark 4.9 concerns two points: 
The well-definedness of the in-plane bending power term, and its parametrization-dependence.

We agree that the internal power of Steigmann, see \eqref{e:workpairGradientTheoryCM}, is not wrong and does not miss any contributions in the present context. 
This was shown above.
Also it does not depend on the surface parameterization (i.e.~the choice of curvilinear coordinate system), as Steigmann rightly points out.\\
However, there is still an issue with the last term in \eqref{e:workpairGradientTheoryCM}:
It can lead to constitutive models for in-plane bending that are not well-defined for some deformations.\\
Steigmann's expression \eqref{e:workpairGradientTheoryCM} leads to the constitutive equations for the membrane and bending stresses,
\eqb{l}
\tau^{\alpha\beta} = 2\ds\pa{W}{a_{\alpha\beta}}\,,\quad
M_0^{\alpha\beta} = \ds\pa{W}{b_{\alpha\beta}}\,,\quad
\Mbarab_{0\gamma} = \ds\pa{W}{S_{\alpha\beta}^\gamma}\,,
\label{e:consti-DS}\eqe
based on the stored energy function $W = W\big(a_{\alpha\beta},b_{\alpha\beta},S_{\alpha\beta}^\gamma\big)$.
Our expression \eqref{e:workpair3} on the other hand, leads to the constitutive equations 
\eqb{l}
\tau^{\alpha\beta} = 2\ds\pa{W}{a_{\alpha\beta}}\,,\quad
M_0^{\alpha\beta} = \ds\pa{W}{b_{\alpha\beta}}\,,\quad
\Mbarab_0 = \ds\pa{W}{\bar b_{\alpha\beta}}\,,
\label{e:consti}\eqe
based on the stored energy function $W = \hat W\big(a_{\alpha\beta},b_{\alpha\beta},\bar b_{\alpha\beta}\big)$.
\\
In general, the in-plane bending behavior of fibers should be related to a change of their curvature $\kappa_\mrg$.
Thus, the in-plane bending energy should be a function of $\kappa_\mrg$.
This means that in general the in-plane bending energy cannot be a function of $S^\gamma_{\alpha\beta}$ alone, as it does not fully describe $\kappa_\mrg$ according to \eqref{e:kappa}.
The problem with the third terms in Eqs.~\eqref{e:workpairGradientTheoryCM} and \eqref{e:consti-DS} is that they suggest precisely that. 
Indeed, such a function has been proposed in \cite{Steigmann2015}, cf.~Eq.~(60), and used in the computational model of \cite{Schulte2020}, cf.~Eq.~(42).
It can lead to spurious bending moments as is shown in the example below.
In our formulation this problem does not appear, since $\kappa_\mrg$ is fully described by $\bar b_{\alpha\beta}$, i.e.~$\kappa_\mrg = \bar b_{\alpha\beta}\,\ell^{\alpha\beta}$, cf.~Eq.~(50) of \cite{shelltextile}. 

\begin{figure}[H]
\begin{center} \unitlength1cm
\begin{picture}(0,4.7)
\put(-4.1,-.07){\includegraphics[height=50mm]{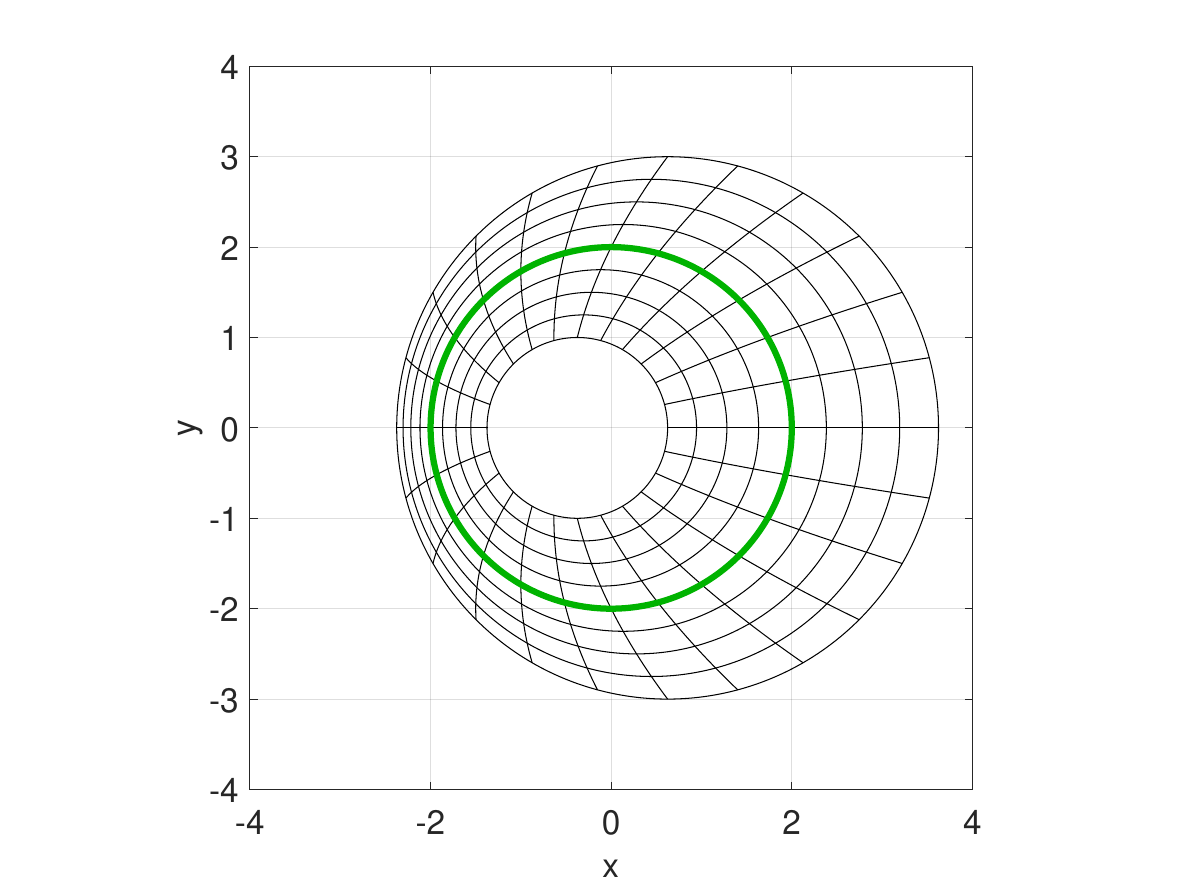}}
\put(-9.35,-.07){\includegraphics[height=50mm]{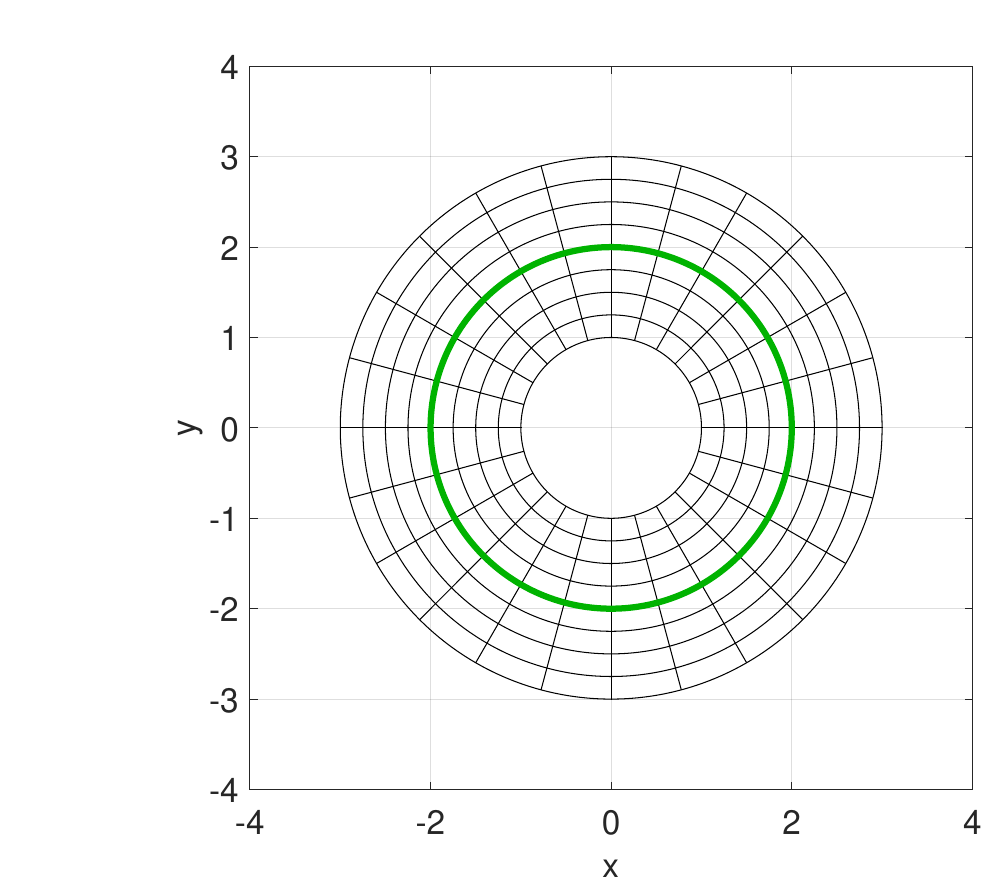}}
\put(2.05,-.15){\includegraphics[height=48mm]{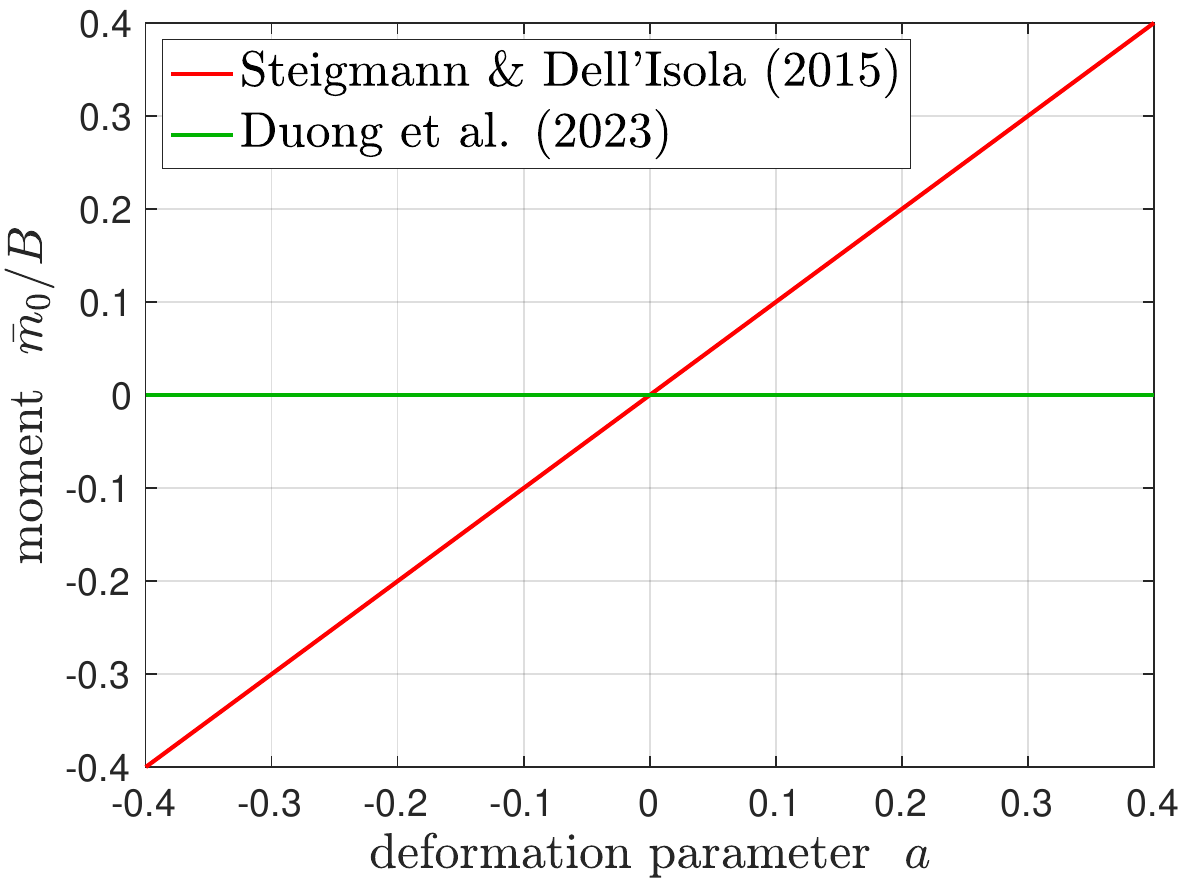}}
\put(-8.45,-.1){{\small{a.}}}
\put(-3.2,-.1){{\small{b.}}}
\put(2.1,-.1){{\small{c.}}}
\end{picture}
\caption{Deformation example of Eqs.~\eqref{e:X} and \eqref{e:u}: a.~Initial configuration for $r_1=1$, $r_0=2$ and $r_2=3$; b.~Deformed configuration for $a_1 = 0.25$ and $a_2 = 0$; c.~Corresponding fiber bending moment $\bar m_0$ according to the constitutive models of \cite{Steigmann2015} and \cite{shelltextile}.
The former predicts an increase of $\bar m_0$ with the deformation, even though the fiber curvature does not change. 
Here $a := a_1\cos\phi+a_2\sin\phi$.\\[-5mm]}
\label{f:ex}
\end{center}
\end{figure}
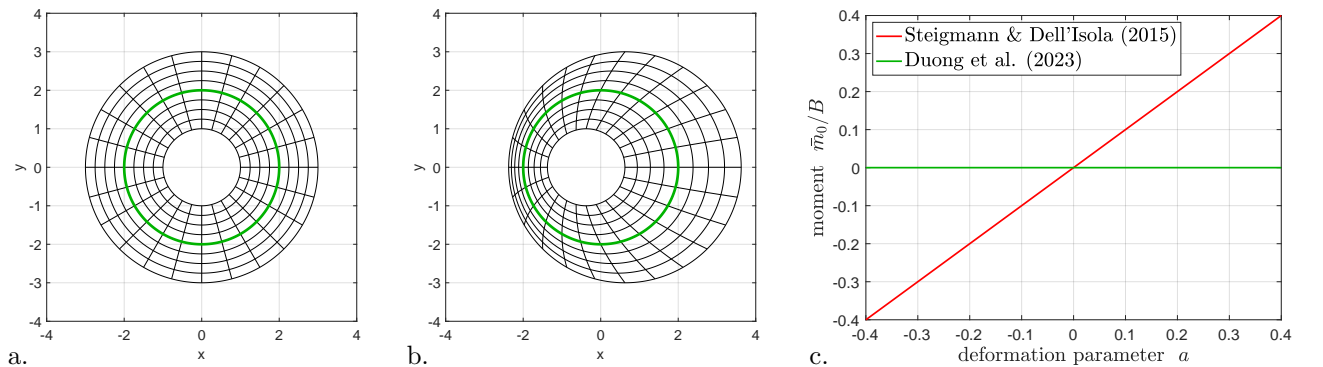

The issue can be illustrated by the following example, see Fig.~\ref{f:ex}.
We consider a ring-shaped domain described by the parameterization
\eqb{l}
\bX(r,\phi) = r\cos\phi\,\be_1 + r\sin\phi\,\be_2\,,\quad
0 \leq \phi < 2\pi\,,\quad
r_1 \leq r\leq r_2\,,
\label{e:X}\eqe
with a circular fiber located at $r = r_0$.
The domain is deformed by applying the displacement field  
\eqb{l}
\bu(r,\phi) = \ds \frac{a_1}{2}\,\big(r^2 - r_0^2\big)\,\be_1 +  \frac{a_2}{2} \,\big(r^2 - r_0^2\big)\,\be_2\,,
\label{e:u}\eqe
where $a_1$ and $a_2$ are parameters of $\bu$ that have sufficiently small magnitude to avoid singularities in the deformation $\bx = \bX + \bu$.
For any admissible $a_i$, displacement $\bu$ is zero along the fiber, and hence does not change its curvature, which remains equal to $1/r_0$.
(The fiber is only stretched in lateral direction.)
Without curvature change, no fiber bending moment should appear.
However, according to the model of \cite{Steigmann2015}, the third term in \eqref{e:consti-DS} generates the fiber bending moment (see Appendix A)
\eqb{l}
\bar m_0 \sim a_1\cos\phi+a_2\sin\phi\,,
\label{e:barm}\eqe
which increases with $a_i$, see Fig.~\ref{f:ex}c.
On the other hand, for the constitutive model proposed in \cite{shelltextile}, no fiber bending moment appears.
The reason for the spurious moment in Eq.~\eqref{e:barm} lies in the constitutive model of \cite{Steigmann2015} that is based on $S_{\alpha\beta}^\gamma$ alone.
As seen in Eq.~\eqref{e:kappa} -- and illustrated by the example in Sec.~7.1 of \cite{shelltextile} -- $S_{\alpha\beta}^\gamma$ yields an incomplete curvature description.
The constitutive model of \cite{Steigmann2015} is a natural choice following from Eq.~\eqref{e:workpairGradientTheoryCM}.
Therefore Eq.~\eqref{e:workpairGradientTheoryCM}, which itself is correct, can suggest spurious bending models. 

As the example shows, the last term in Eq.~\eqref{e:consti-DS} can generate a fiber moment that depends inconsistently on the deformation, seen here through the surface deformation parameters $a_i$.
However, the third term in Eq.~\eqref{e:workpairGradientTheoryCM} does not depend on the surface parametrization, 
as was written imprecisely in Remark~4.9.

\subsubsection*{Concerning the remaining points raised by Steigmann}{\label{s:other}}

Steigmann claims that our ``force and torque balance laws are postulated on the basis of free-body diagrams without reference to the Principle of Virtual Power".
This is not true.
Our formulation constructs the force and torque balance laws and the constitutive equations systematically from linear momentum balance, angular momentum balance, and the mechanical power balance in Secs.~3.3, 3.4 and Sec 3.5 of \cite{shelltextile}. 
The latter balance is mathematically equivalent to the principle of virtual power.

Steigmann further states that our formulation does ``not make explicit the famous Kirchhoff boundary conditions and corner forces established by Kirchhoff".
Such corner forces are fully contained in our formulation. 
They appear when the boundary traction $\bT$ is replaced by its effective counterpart and the domain boundary of the shell surface is not smooth, which is shown in Appendix B below.
The corner forces are part of the external virtual work, and so the internal virtual work expression presented in \cite{shelltextile} is unaffected.

Singular boundary fibers are indeed not studied in our work.
But they do not affect the examples presented in \cite{shelltextile} and \cite{textshell2}.

\subsubsection*{Appendix A: Derivation of the bending moments in the example of Fig.~\ref{f:ex}}

Introducing $c:=\cos\phi$ and $s:=\sin\phi$, the deformation defined by Eqs.~\eqref{e:X} and \eqref{e:u} has the tangent vectors along parameters $r$ and $\phi$,
\eqb{l}
\ba_1 = \bx_{,r} = (c+a_1\,r)\,\be_1 + (s+a_2\,r)\,\be_2\,,\quad
\ba_2 = \bx_{,\phi} = (-s\,\be_1+c\,\be_2)\,r\,,
\eqe
and their derivatives
\eqb{l}
\ba_{1,1} = a_1\,\be_1 + a_2\,\be_2\,,\quad
\ba_{1,2} = -s\,\be_1+c\,\be_2\,,\quad
\ba_{2,2} = -(c\,\be_1+s\,\be_2)\,r\,.
\eqe
From this follows
\eqb{l}
[a_{\alpha\beta}] = \left[\begin{array}{cc}
(c+a_1\,r)^2 + (s+a_2\,r)^2  & (a_2\,c - a_1\,s)\,r^2 \\
 (a_2\,c - a_1\,s)\,r^2 & r^2
\end{array}\right],
\eqe
\eqb{lll}
\det[a_{\alpha\beta}] = g^2\,r^2\,,\quad
g := 1 + (a_1\,c +  a_2\,s)\,r
\eqe
and
\eqb{l}
\big[a^{\alpha\beta}\big] = \ds\frac{1}{g^2\,r^2} \left[\begin{array}{cc}
r^2  & (a_1\,s-a_2\,c)\,r^2 \\
 (a_1\,s-a_2\,c)\,r^2 & (c+a_1\,r)^2 + (s+a_2\,r)^2
\end{array}\right],
\eqe
which leads to the dual tangent vectors
\eqb{l}
\ba^1 = \ds\frac{c\,\be_1 + s\,\be_2}{g}\,,\quad
\ba^2 = \ds\frac{- (s+a_2\,r)\,\be_1 + (c+a_1\,r)\,\be_2}{g\,r}\,.
\eqe
The Christoffel symbols thus become
\eqb{l}
\big[\Gamma^1_{\alpha\beta}\big] = \ds\frac{1}{g}\left[\begin{array}{cc}
a_1\,c + a_2\,s & 0 \\
0 & -r
\end{array}\right], \quad
\big[\Gamma^2_{\alpha\beta}\big] = \ds\frac{1}{g\,r}\left[\begin{array}{cc}
a_2\,c-a_1\,s & g \\
g & (a_2\,c-a_1\,s)\,r^2 
\end{array}\right].
\eqe
The corresponding quantities for the reference configuration ($\bA_\alpha$, $\bA_{\alpha,\beta}$, $A_{\alpha\beta}$, $A^{\alpha\beta}$, $\bA^\alpha$ and $\bar\Gamma^\gamma_{\alpha\beta}$) follow from this with $a_1=a_2=0$ and $g = 1$.
Thus
\eqb{lllll}
\big[S^1_{\alpha\beta}\big] \dis \big[\Gamma^1_{\alpha\beta} -\bar\Gamma^1_{\alpha\beta}\big] 
\is \ds\frac{a_1\,c + a_2\,s}{g}\left[\begin{array}{cc}
1 & 0 \\
0 & r^2
\end{array}\right], \\[5mm]
\big[S^2_{\alpha\beta}\big] \dis \big[\Gamma^2_{\alpha\beta} -\bar\Gamma^2_{\alpha\beta}\big] 
\is \ds\frac{a_2\,c-a_1\,s}{g\,r}\left[\begin{array}{cc}
1 & 0 \\
0 & r^2
\end{array}\right],
\label{e:S}\eqe
which is only zero for vanishing $a_i$.
Further $J = \sqrt{\det[a_{\alpha\beta}]/\det[A_{\alpha\beta}]} = g$.
\\
For the fiber at $r = r_0$, the fiber direction (before and after deformation) is
\eqb{lll}
\bL =  \bell = \ds\frac{\ba_2}{\norm{\ba_2}} =  -s\,\be_1 + c\,\be_2\,,
\eqe
such that
\eqb{l}
L^1 = \ell^1 := \bell\cdot\ba^1 = 0\,,\quad
L^2 = \ell^2 := \bell\cdot\ba^2 = \ds\frac{1}{r_0}\,.
\label{e:La}\eqe
With this, the curvature-invariant of \cite{Steigmann2015} follows as
\eqb{lll}
\bg_\mrL := L^\alpha\,L^\beta\,S_{\alpha\beta}^\gamma\,\ba_\gamma = a_1\,\be_1 + a_2\,\be_2\,,
\eqe
such that fiber bending energy of \cite{Steigmann2015}, 
\eqb{l}
W_\mathrm{SD} := \ds\frac{B}{2}\,\bg_{\mrL}\cdot\bg_{\mrL} = \ds\frac{B}{2}\big(a_1^2 + a_2^2\big)\,,
\label{e:WSD}\eqe
increases with $a_1$ and $a_2$.
Here, $B$ is the material constant for in-plane bending.
\\
The fiber director follows from $\bell$ by a counterclockwise rotation of $90^\circ$, i.e.
\eqb{lll}
\bc = -c\,\be_1 - s\,\be_2\,.
\eqe
Thus
\eqb{l}
c_1 := \bc\cdot\ba_1 = g_0 := g|_{r=r_0}\,,\quad
c_2 := \bc\cdot\ba_2 = 0\,.
\label{e:ca}\eqe
With this and $\bL_{,2} = -c\,\be_1-s\,\be_2$, the two contributions of the geodesic curvature in \eqref{e:kappa} become
\eqb{l}
\kappa_\mrg^\Gamma := \ell^\alpha\,\ell^\beta\,S_{\alpha\beta}^\gamma\,c_\gamma = -a_1\,c - a_2\,s\,,
\label{e:kG}
\eqe
and
\eqb{l}
\kappa_\mrg^\mrL
:= \lambda^{-1}\,c_\alpha\,\ell^\beta\,\bA^\alpha\cdot\bL_{,\beta}  = c_1\,\ell^2\,\bA^1\cdot\bL_{,2} 
= \ds\frac{1}{r_0} + a_1\,c + a_2\,s\,,
\eqe
for the present case, where there is no fiber stretch ($\lambda=1$).
The full geodesic curvature thus is
\eqb{l}
\kappa_\mrg = \kappa_\mrg^\Gamma + \kappa_\mrg^\mrL =  \ds\frac{1}{r_0}\,,
\label{e:ktot}
\eqe
which is the expected value for the example.
This is equal to the geodesic curvature in the reference configuration (i.e.~$\kappa_\mrg^0 = 1/r_0$).
All the relative curvature measures of \cite{shelltextile} are thus zero, and the in-plane bending energy of \cite{shelltextile},
\eqb{l}
W = \ds\frac{B}{2}\big(\kappa_\mrg-\kappa_\mrg^0\big)^2\,,
\eqe 
vanishes for all $a_i$.
In our formulation the in-plane bending moment in the fiber is
\eqb{l}
\bar m_0 = \ds\pa{W}{\kappa_\mrg}\,,
\label{e:m0}\eqe
which gives $\bar m_0 = B\big(\kappa_\mrg-\kappa_\mrg^0\big)=0$, here.
Eq.~\eqref{e:m0} follows from Eqs.~(82b), (108c) and (110c) of \cite{shelltextile} with $\bnu=\bell$ and $\bar m_0 = J\,\bar m$.
\\
The corresponding (equivalent) bending moment according to formulation \eqref{e:workpairGradientTheoryCM} is 
\eqb{l}
\bar m_0 = \bar M^{\alpha\beta}_{0\,\gamma}\,\ell_\alpha\,\ell_\beta\,c^\gamma\,.
\label{e:m0M}\eqe
This follows from inserting \eqref{e:var_kapp} and $\bar\mu_0 = \bar m_0$ into \eqref{e:workpair1} and comparing corresponding terms with \eqref{e:workpairGradientTheoryCM}. 
Here $\ell^\alpha \ell_\alpha = 1$ and $c^\alpha c_\alpha = 1$ have been used.
For the constitutive model of \cite{Steigmann2015} follows
\eqb{l}
\bar M^{\alpha\beta}_{0\,\gamma} = B\,L^{\alpha\beta}\,L^{\mu\eta}\,S^\delta_{\mu\eta}\,a_{\gamma\delta}
\eqe
from Eqs.~(\ref{e:consti-DS}c) and \eqref{e:WSD}.
Eq.~\eqref{e:m0M} then yields
\eqb{l}
\bar m_0^\mathrm{SD} = B\,\lambda^2\,S^\gamma_{\alpha\beta}\,L^{\alpha}\,L^{\beta}\,c_{\gamma}\,,
\eqe
for general fibers with
$L^\alpha = \lambda\,\ell^\alpha$.
With Eqs.~\eqref{e:S}, \eqref{e:La} and \eqref{e:ca} of 
the present example, this becomes $\bar m_0^\mathrm{SD} = B\,(a_1\,c+a_2\,s)$, leading directly to Eq.~\eqref{e:barm}.

\subsubsection*{Appendix B: Corner forces in our formulation}

Corner forces appear when the external virtual work expression (here for a single fiber family)
\eqb{lll}
G_\mathrm{ext} \is \ds\int_{\sS}\delta\bx\cdot\bff\,\dif a 
+ \ds\int_{{\partial\sS}} \delta\bx\cdot\bT\,\dif s + \ds \int_{\partial\sS} \delta\bn\cdot\bM\,\dif s + \ds  \int_{\partial\sS} \delta\bc\cdot\bMbar\,\dif s
\label{e:Giie}
\eqe
(cf.~Eq.~(135.3) in \cite{shelltextile}) is rewritten following the procedure used in our previous work (\cite{shelltheo},  cf.~Sec 6.3).
This starts with the balance of angular momentum
\eqb{l}
\ds \frac{D}{Dt} \int_{{\sR}}\rho\,\bx\times\bv\, \dif a  =  \int_{{\sR}}\bx\times\bff\, \dif a + \ds \int_{\partial{\sR}} \bx\times\bT\, \dif s +   \ds \int_{\partial{\sR}} {\bmhat}\, \dif s~,
\label{e:angmomentglobal}
\eqe
(cf.~Eq.~(87) in \cite{shelltextile}), where ${\bmhat}:=m_\tau\,\btau + m_\nu\,\bnu + \mbar\,\bn$ is the complete boundary bending moment defined in Eq.~(66) in \cite{shelltextile}. 
Using $\bnu = \btau\times\bn$, $\bn = -\btau\times\bnu$ and $\btau:=\partial\bx/\partial s = \bx'$, the last two terms in Eq.~\eqref{e:angmomentglobal} can be rewritten as
\eqb{llllll}
\bx \times\bT \,+\, \bmhat
\is \bx \times\bT \,+\, m_\nu\,\bx'\times\bn  \,-\, \mbar\,\bx'\times\bnu \,+\, m_\tau\,\btau \\[1.5mm]
\is \bx \times\bT \,+\, (m_\nu\,\bx\times\bn)' \,-\, \bx\times (m_\nu\,\bn)'  \,-\, (\mbar\,\bx\times\bnu)'  \,+\, \bx\times (\mbar\,\bnu)' \,+\, m_\tau\,\btau\\[1.5mm]
\is \bx\times\Big[\bT - (m_\nu\,\bn)' + (\mbar\,\bnu)'\Big] \,+\, m_\tau\,\btau \,+\, (m_\nu\,\bx\times\bn)' \,-\, (\mbar\,\bx\times\bnu)' \,.
\eqe
The last expression shows that the moment components $m_\nu$ and $\mbar$ contribute to the effective boundary traction
\eqb{l}
\hat\bt := \bT - (m_\nu\,\bn)' + (\mbar\,\bnu)'\,.
\label{e:effT}
\eqe
With this, $\delta\bx'\cdot\bn=-\btau\cdot\delta\bn$ and $\delta\bx'= \tau^\alpha\,\delta\ba_\alpha$ (following from $\btau\cdot\bn=0$ and $\bx'=\tau^\alpha\ba_\alpha$ with $\tau^\alpha := \partial\xi^\alpha/\partial s$ being fixed), we can write
\eqb{l}
\delta\bx\cdot\bT = \delta\bx\cdot\hat\bt \,+\, \delta\bn\cdot m_\nu\,\btau \,+\, \delta\ba_\alpha\cdot\mbar\,\tau^\alpha\,\bnu \,+\, (m_\nu\,\delta\bx\cdot\bn)' \,-\, (\mbar\,\delta\bx\cdot\bnu)'.
\label{e:deltaT}
\eqe
Inserting Eq.~\eqref{e:deltaT} into Eq.~\eqref{e:Giie} and using $\bM = m_\tau\,\bnu - m_\nu\,\btau$ (cf.~Eq.~(58) from \cite{shelltheo}),
$\bar\bM = -\bar m\,\bell$ and $\delta\bc\cdot\bell=-\ell^\alpha\bc\cdot\delta\ba_\alpha$ (cf.~Eqs.~(76c) and (212) from \cite{shelltextile})
gives
\eqb{lll}
G_\mathrm{ext} \is \ds\int_{\sS}\delta\bx\cdot\bff\,\dif a 
+ \ds\int_{{\partial\sS}} \delta\bx\cdot\hat\bt\,\dif s + \ds \int_{\partial\sS} \delta\bn\cdot m_\tau\,\bnu \,\dif s 
+ \ds\int_{\partial\sS} \delta\ba_\alpha\cdot\mbar\,\big(\tau^\alpha\bnu + \ell^\alpha\,\bc\big)\,\dif s  \\[4mm]
\plus \Big[\delta\bx\cdot m_\nu\,\bn\Big] -  \Big[\delta\bx\cdot\mbar\,\bnu\Big]\,,
\label{e:Giie2}
\eqe
where the last two terms are jump terms that appear at corners on $\partial\sS$ due to the out-of-plane boundary moment $m_\nu$ and the in-plane boundary moment $\bar m$. 
Effectively, these moments act as point forces at those corners.
Contrary to $m_\nu$, $\bar m$ also contributes to another term in $G_\mathrm{ext}$ -- the fourth term in Eq.~\eqref{e:Giie2}. 
This is also seen in the formulation of \cite{Steigmann2018}, cf.~Eq.~(87).

\bigskip

\bibliographystyle{apalike}
\bibliography{sauerduong}

\end{document}